# All-optical probe of precessional magnetization dynamics in exchange biased NiFe/FeMn bilayers

M.C. Weber[1],* , H. Nembach[1], B. Hillebrands[1], and J. Fassbender[2]

[1]*Fachbereich Physik and Forschungsschwerpunkt MINAS, Technische Universität Kaiserslautern, Erwin-Schrödinger-Straße 56, 67663 Kaiserslautern, Germany*

[2]*Institut für Ionenstrahlphysik und Materialforschung, Forschungszentrum Rossendorf, 01314 Dresden, Germany*

**Abstract**

An internal anisotropy pulse field is launched by an 8.3 ps short laser excitation, which triggers precessional magnetization dynamics of a polycrystalline NiFe/FeMn exchange bias system on the picosecond timescale. Due to the excitation the unidirectional anisotropy and, thus, the exchange coupling across the interface between the ferromagnetic and the antiferromagnetic layer is reduced, leading to a fast reduction of the exchange bias field and to a dramatic increase of the zero-field susceptibility. The fast optical unpinning is followed by a slower recovery of the interfacial exchange coupling dominated by spin-lattice and heat flow relaxation with a time constant of the order of 160 ps. The measured picosecond time evolution of the exchange decoupling and restoration is interpreted as an anisotropy pulse field giving rise to fast precessional magnetization dynamics of the ferromagnetic layer. The strength of the internal pulse field and even the initial magnetization deflection direction from the equilibrium orientation can be controlled by the absorbed photons. The dependence of the effective Gilbert damping on both small and large angle precessional motion was studied, yielding that both cases can be modeled with reasonable accuracy within the Landau-Lifshitz and Gilbert framework.

* Electronic mail: mweber@physik.uni-kl.de

## I. Introduction

As data transfer rates increase in magnetic recording, the high frequency performance of related magnetic devices becomes increasingly important. Controlling magnetization dynamics requires the understanding and tuning of relaxation mechanisms, and reliable techniques are needed to observe such processes at high frequencies deep within the GHz regime.

It has been shown that short photoexcitations of a magnetic medium can create hot, non-equilibrium spins [1-3]. Thus, the magnetic response is modulated at a microscopic level on a picosecond or even subpicosecond time scale. Internal anisotropy pulse fields with rise times of the order of the exciting laser pulse width can be induced, which trigger ultrafast coherent precessional dynamics in both ferromagnetic layers and exchange coupled bilayers [4,5]. Even fast spin reorientation processes in antiferromagnets have been excited [6]. The optical approach raises the prospect of inducing and investigating fundamental spin dynamics and relaxation phenomena in the time domain with the system driven by the absorbed photons. As a specific material test system we propose an exchange bias system, which is inherently related to the spin temperature and where local fluctuations of the interfacial exchange coupling might represent an additional damping channel [7,8].

A problem of contemporary interest at both fundamental and applied levels concerns a deeper understanding of the origin of the unidirectional exchange bias anisotropy in bilayer systems consisting of a ferromagnetic (F) and an adjacent antiferromagnetic (AF) layer. This unidirectional anisotropy is found, for instance, when the F-AF-bilayer system is cooled below the Néel temperature of the AF layer in an applied magnetic field. The uncompensated AF spins are aligned with respect to the F layer and frozen. Thus, the interfacial AF-spins are the origin of an additional internal magnetic field, the so-called exchange bias field $H_{eb}$. This results in a shift of the ferromagnetic hysteresis loop with respect to zero-applied field. Due to the exchange coupling, the easy axis coercivity is additionally enlarged compared to a single F layer of the respective material. The potential to pin a magnetization direction leads to a variety of applications which rely on spin-valve systems, ranging from magnetic random access memories (MRAM), hard disk read heads exploiting magnetoresistive effects, to magnetic position sensors and galvanic isolators. In each case, a magnetic reference direction is required which is established by the exchange bias effect. Considerable insight into the exchange bias effect has been gained from theoretical micromagnetic modeling and quasistatic magnetization reversal studies, indicating that the microstructure of the heterointerface plays a key role in such exchange coupled systems [9-11]. For recent reviews of the exchange bias effect, see Refs. 12 and 13.

Exploiting the possibility of real time measurements of magnetization precession upon photoexcitation, the magnetic damping behaviour described by the phenomenological Gilbert damping parameter $\alpha$ can be investigated. Recently, Silva *et al.* [14] have shown that nonlinear effects can play a crucial role in the large angle precession dynamics regime. Even more recently, Nibarger *et al.* have demonstrated that both the large as well as the small angle free precession magnetization response to pulsed field excitations can be modeled with reasonable accuracy in terms of Landau-Lifshitz and Gilbert (LLG) damping [15]. At the same time, the good fits of LLG damping to large angle precession data indicate that there are no nonlinear damping effects present in these material systems. Many years ago, moreover, Patton *et al.* showed that the large angle precession response associated with domain wall motion in Permalloy films could also be modeled through the Landau-Lifshitz and Gilbert equation with a relatively small value of the Gilbert damping parameter [16].

In the present article we address the dynamics of the exchange coupling in exchange bias systems with special emphasis on precessional magnetization dynamics due to laser excitations. In our case, the photoexcitation is on the low picosecond timescale, i.e., the temperature associated with the spin and electron subsystems is in equilibrium already. Vaterlaus *et al.* have shown that typical spin-lattice relaxation times of ferromagnets are less than 100 ps [17]. The investigated magnetization dynamics is therefore related to spin-lattice relaxation dynamics on the one hand and fast local heating of the phonon subsystem on the other hand, covering a timescale from below 100 ps up to several nanoseconds. An ideal test system for this purpose is an exchange bias bilayer system, since the exchange bias effect is known to depend strongly on temperature. Hence, the anisotropy in the system originating from the exchange bias effect can be switched on and off by a change in spin temperature. Pioneering work in this field has been performed by Ju *et al.* [4] who examined the NiFe/NiO exchange bias system by means of all-optical femtosecond pump-probe experiments.

Our aim is to investigate the time constants involved in the fast unpinning and recovery of the exchange coupling across the interface in the metallic NiFe/FeMn exchange bias system and to study the dynamic response of the F layer due to the photomodulation of the exchange bias anisotropy within the GHz regime. The dependence of the effective Gilbert damping parameter on the strength of the internal pulse field, i.e., the angle of precession is studied in detail. This work may open the way to a picosecond photocontrol of the free layer magnetization in spin-valve systems and holds the potential for heat-induced magnetization rotation or even heat-assisted magnetic recording [18] with temperatures involved far below the Curie temperatures of single ferromagnetic films. The optical approach raises the prospect of investigating the basic magnetization dynamics and relaxation phenomena in real time, which allows for a direct comparison with solutions of the LLG equation.

## II. Experiment

The polycrystalline exchange bias samples have been prepared by UHV evaporation using both electron beam and effusion cell evaporation. As a growth template a 15 nm thick Cu buffer layer on top of a thermally oxidized Si substrate has been used. The exchange bias system itself consists of a 5 nm thick $Ni_{81}Fe_{19}$ (F) and a 10 nm thick $Fe_{50}Mn_{50}$ (AF) layer. Finally, the sample was covered with a 2 nm Cr cap layer to protect it from oxidation. In order to initialize the exchange bias effect, the sample was field cooled from above the blocking temperature $T_B$ (155 °C) after deposition. Temperature dependent quasistatic Kerr magnetometry has been used to check for the unidirectional anisotropy character and the blocking temperature of the samples. For further details, see Ref. 19.

The stroboscopic all-optical pump-probe experiments have been performed employing 8 ps short laser pulses generated by a SESAM mode-locked diode pumped Nd:YVO$_4$ oscillator running at a repetition rate of 80 MHz and a wavelength of 1064 nm [20]. After amplification, the laser beam is inserted into a second harmonic unit which maintains the pulse duration and delivers laser pulses of a 532 nm wavelength and a maximum pulse energy of 50 nJ. The beam is then divided into an intense pump beam, with an energy of 11 nJ per pulse for the presented results, and a weak probe beam by a beamsplitter. The intense pump pulse is directed nearly normal to the sample surface and focused to a spot diameter of about 30 μm yielding a power density of about 0.3 GW/cm$^2$. The polarization of the pump beam is rotated 90° with respect to the polarization of the probe beam avoiding an undesired crosstalk between the reflected probe beam and scattered pump light at the detection side. The probe pulse is time delayed by a motorized translation stage covering a time interval from a negative delay of -400 ps (probe pulse is arriving before the pump pulse on the sample) up to about 6 ns. The overlap between pump and probe spot has been carefully defined, and it has been checked that there is no influence of the weak probe pulse on the observed transient magnetic phenomena. Time resolved magneto-optics as a reliable probe of the magnetization is employed. Thus, the probe beam is used to sense the optically induced changes of the magnetization of the NiFe layer by means of longitudinal magneto-optic Kerr effect (spot diameter: 25 μm).

In order to investigate the optical control of the magnetization of the ferromagnetic layer of the exchange coupled bilayer in the time domain, the quasistatic hysteresis loop is sensed by a probe pulse with a fixed time delay to the laser excitation pulse. The quasistatic hysteresis loop then reflects the magnetic parameters present for a given time delay. The hysteresis loops are recorded with a field sweep rate of 1 Oe/s to suppress the influence of thermal activated training on the left and right side coercive fields, i.e., on the shift field of the easy axis hysteresis loops during recording [21]. Hence, only pump pulse induced effects are observed. Real time measurements of precessional magnetization dynamics of the F layer have been performed using stroboscopic transient Kerr magnetometry by keeping the applied magnetic field constant, thus, defining the initial magnetization state and varying the time delay between the pump and probe pulse continuously. The proportionality of the longitudinal Kerr effect to the in-plane magnetization component makes the transient experiment also sensitive to photoinduced changes in the direction of the magnetization, e.g., due to coherent rotation.

### III. Results

Since the magnitude of the exchange bias field is inherently related to temperature, the pump laser pulse is expected to serve as a local "heating" pulse influencing the exchange coupling across the heterointerface. Thereby, the exchange bias field is manipulated. Since the pump pulse duration and the time resolution of our system is on the low picosecond timescale, only spin-lattice and predominantly lattice heating effects can be observed. Additional experiments using fs-laser pulses addressing the thermalization of the heated electron and spin system for these kind of samples are currently set up for further investigations.

First, the transient hysteresis loops along the easy magnetization direction are investigated using the above described stroboscopic measurement scheme. The measurements have been performed in a time window ranging from a negative time delay of -370 ps between pump and probe pulse, i.e., the probe pulse arrives 12.13 ns after the pump pulse, up to about 3000 ps after the arrival of the pump pulse. Details of the measurement can be found elsewhere [22]. Within this time range all time-dependent effects are already relaxed (see Fig. 1). For each delay time the shift field was evaluated according to $H_{eb} = \left( H_{c,left} + H_{c,right} \right)/2$, with $H_{c,left}$ and $H_{c,right}$ representing the left and right coercive fields of the loop, respectively. Hence, figure 1 yields the time evolution of the shift field $H_{eb}(t)$. No pump induced effects on the shift field can be observed for negative time delays compared to an easy axis hysteresis loop recorded with a blocked pump beam. An initial exchange bias field of $H_{eb,init}$ = -123 Oe and a coercive field of $H_{c,init}$ = 24 Oe are observed. For a positive delay time of 10 ps a maximum reduction of the exchange bias field to a value $H_{eb}$ = -65 Oe is observed, which is equivalent to a reduction of about 47 % (see also Fig. 3a). Furthermore, analyzing the time evolution of the easy axis coercive field, we observe a reduction of $H_c$ within the first 10 ps as well (see open squares in Fig. 1). For larger positive time delays a slower recovery of the exchange bias field as well as the coercivity on a picosecond timescale can clearly be observed.

In order to analyze the time evolution of both the shift $H_{eb}$ and the coercive field $H_c$ more quantitatively, the data have been fitted to a phenomenological model introduced by Ju *et al.* [4,22] based on an extension of a thermal activation model [23,24] to the picosecond timescale. This model describes the process of photomodulation by a time-dependent single exponential driving term,

$$H_x(t) = H_{x,\text{init}}\left(1 - m \cdot \exp\left(-t/\tau\right)\right), \quad \text{with } x = eb, c \tag{1}$$

where $H_{eb,\text{init}}$, $H_{c,\text{init}}$ describe the initial bias and coercive field value respectively. $m$ is the so called modulation depth, i.e., the strength of the photomodulation, whereas $\tau$ represents the time constant of the recovery of both the exchange bias and coercive field. The internal relaxation time $\tau$ upon photoexcitation comprises both an interfacial spin-lattice relaxation and heat flow dynamics. The best fit to our experimental values is shown as a full line in Fig. 1 with $m = 0.48$ and $\tau = 160$ ps. A fit to the measured time evolution of the coercive field $H_c$ yields a consistent internal relaxation time of 160 ps upon photoexcitation but a smaller modulation depth of $m = 0.42$ (see open squares in Fig. 1), indirectly supporting the idea that two different mechanisms contribute to the shift of the hysteresis loop and an enlarged easy axis coercivity in exchange biased bilayers [25].

Next, the time evolution of the zero-field susceptibility of the hard axis magnetization reversal loops is investigated. It is addressed, whether the time dependence matches the possible exchange coupling modulation found in the easy axis case. Hard axis reversal loops are recorded for delay times from -250 ps up to about 2000 ps. Details are reported elsewhere [22]. The magnetization reversal loops taken at a negative delay time show the expected hard axis behavior for exchange bias systems. In order to analyze these data more quantitatively, the zero-field susceptibility has been extracted from the magnetization reversal curves by performing a linear fit to the data close to zero-applied field. The slope corresponds to the zero-field susceptibility $\chi$, which is plotted in Fig. 2 as a function of the pump-probe delay time. Shortly after the arrival of the excitation pulse, a dramatic change in the hard axis loop shape is observed leading to a sharp increase of $\chi$. 10 ps after the heating pulse, the loop shape resembles the shape of a NiFe film easy axis loop. Hence, it reflects a thermally induced partial decoupling of the F layer from the AF layer (see also Fig. 5a). The reversal of the F layer magnetization becomes quasi-isotropic. Figure 2 shows the measurement around 0 ps pump-probe delay together with the measured background-free (non-collinear) intensity autocorrelation of the pump pulse at the sample position (solid line), yielding a measure of the temporal pump pulse width. It clearly demonstrates that the partial collapse of the interfacial exchange coupling appears well within the pump pulse width. For larger delay times, the hard axis behavior starts to be restored, thus, the susceptibility $\chi$ reaches its initial value.

The internal time constant involved in the hard axis geometry is extracted by a fit to Eq. (2) with a modified phenomenological modulation depth $m^*$ and a recovery time $\tau^*$

$$\chi(t) = \chi_{\text{init}} \left( 1 + m^* \cdot \exp(-t/\tau^*) \right) \qquad (2)$$

The best fit is achieved with the values of $\tau^* = 160$ ps and $m^* = 13.75$ (dashed-dotted line in Fig. 2). The internal relaxation time for both easy ($\tau$) and hard axis ($\tau^*$) magnetization reversal are in good agreement indicating that both time dependencies rely on the same fundamental mechanism of fast unpinning and recovery of the interfacial exchange coupling.

Now, the question arises whether the measured time dependence of the exchange coupling in terms of an effective internal anisotropy pulse field can trigger ultrafast magnetization dynamics of the F layer. The measured time evolutions can be regarded as fingerprints of such an internal excitation field. Even the rise (< 10 ps) and fall (160 ps) times of this transient effective field can be extracted. First, we have a look again at an easy axis hysteresis loop sensed at negative time delay reflecting the situation without pump induced effects and a loop shortly after the arrival of the "heating" pulse (see Fig. 3a) recorded in the above described experiment. If one considers the case of a static applied field of about $H_{\text{stat}} = 125$ Oe, marked by a black dot, two main features are present. First, due to the distribution of exchange bias field directions in the studied polycrystalline bilayer there is already a slight in-plane rotation of the F layer magnetization away from its initial equilibrium orientation (see light grey arrow in Fig. 3a), making the orientation of the magnetization even more sensitive to a change in the spin temperature $T$. Thus, one can expect a maximum change of the equilibrium orientation of the F layer magnetization upon a sudden temperature rise, i.e., a maximum deflection amplitude, for the case of $H_{eb}(T) + H_c(T) < H_{stat} \leq H_{eb,init} + H_{c,init}$ prior to the pump pulse [4]. The photomodulation of the exchange bias shift field gives rise to an internal pulse field, i.e., a transient effective field arises. Moreover, within the described field range the magnetization of the ferromagnetic layer is allowed to change its equilibrium direction upon arrival of the pump pulse, which can induce magnetization dynamics of the ferromagnetic layer of the exchange biased bilayer. Due to the excess energy of the spin system upon an elevated spin temperature, a precession of the magnetization of the ferromagnetic layer about the equilibrium orientation can be induced. Therefore, the marked field value in Fig. 3a represents the most temperature sensitive working point, which allows for a large angle deflection of the magnetization away from its initial equilibrium orientation upon photoexcitation. Furthermore, the thick solid arrow and thick dashed lines in Fig 3a and Fig. 3b enables us to precisely quantify the deflection magnitude in terms of Kerr rotation, which goes beyond the previously reported data [8].

The dynamic response of the F layer magnetization for the above discussed scenario is sensed by transient Kerr measurements. A time resolved Kerr trace is recorded for the case of $H_{stat} =$ 125 Oe, thus, setting the temperature sensitive initial state and varying the pump-probe delay continuously (see Fig. 3b). Indeed, a large-angle precessional oscillation of the magnetization can be resolved with a frequency of about 3.6 GHz. In addition, by choosing different static field values, e.g., different temperature sensitive conditions, we are able to define different internal pulse field strengths (see open dots in 4a). The closer we choose the applied static field to $H_{stat} = 125$ Oe the more pronounced the observed easy axis magnetization oscillations become, which can be understood in terms of an increasing deflection strength on the F layer magnetization (see Fig. 4b). All traces reveal the same fall time of about 150 ps, reflecting a spin-lattice non-equilibrium upon photoexcitiation, and field-dependent oscillation frequencies up to 3.6 GHz. The real time measurements show that the photoinduced changes of the orientation of the magnetization are fully reversible for the applied field values (see Fig. 4b). It is evident that for applied fields that equal the saturation field one cannot induce a precessional response of the magnetization of ferromagnetic layer. In these cases there is no energetic need for the F layer magnetization to change its equilibrium orientation. The only limitation for reversible changes is the pump beam intensity.

In Fig. 5a we exemplarily present reversal loops for the hard axis case, again with (delay 10ps) and without (delay -370 ps) pump induced changes. The transition from a typical hard axis reversal behaviour to a nearly isotropic reversal behaviour can clearly be identified, as already discussed in the analysis of the zero-field susceptibility. In the hard axis case the same argumentation suggesting an internal anisotropy pulse field holds, however, with the additional important feature of a possible control of the direction of the F layer magnetization deflection from the initial orientation upon laser excitation. As suggested by the dashed and full arrows in Fig. 5a, for negative applied field values, i.e., defining the initial state prior to the pump pulse arrival, the initial deflection upon the internal pulse field should go "up", whereas for positive applied fields the deflection should go "down". Figure 5b summarizes the time-resolved hard axis Kerr measurements, providing clear evidence of optically controlled magnetization oscillations and the possibility of controlling the initial deflection orientation. All real time measured traces are offset for clarity. As expected, the maximum oscillation amplitude is found for the traces recorded for applied fields of $H_{stat}$ = +31 Oe and $H_{stat}$ = -31 Oe, respectively (see also solid arrows in Fig. 5a). Again, there is an initial rise and fall time of about 150 to 170 ps, presumably due to the spin-lattice non-equilibrium upon photoexcitation. Furthermore, a field-dependent oscillation period can be identified, as sketched by the light grey dashed arrows in Fig. 5b. The oscillation frequencies $f$ extracted by a fast Fourier transformation (FFT) of the transient Kerr traces for positive field values are plotted as function of the applied field in Fig. 6. A Kittel like square root behaviour ($f \propto \sqrt{H_{stat}}$) is deduced, thus, indicating the homogeneous ($|\vec{k}| = 0$) precessional character of the measured oscillation mode of the ferromagnetic layer of the exchange coupled bilayer. The hard axis precession frequencies turn out to be quite comparable with the measured easy axis precession frequencies. Both time-resolved measurements in easy and hard axis geometry give clear evidence of a strong internal anisotropy pulse field which can externally be controlled by the absorbed pump photons.

## IV. Discussion and Outlook

The experimental observation of the fast reduction of the exchange bias and easy axis coercive fields and the found strong increase of the susceptibility on a 10 ps timescale can be understood in terms of a fast optical unpinning of the exchange coupling at the F/AF heterointerface upon arrival of the pump laser pulse. Non-equilibrium conditions are achieved such that within the laser pulse duration the temperature of the spin and lattice systems is elevated close to the blocking temperature of the exchange bias system, which in turn leads to a collapse of the exchange coupling. Using temperature dependent quasistatic Kerr measurements and extracting the shift field $H_{eb}$ as a function of sample temperature, hence, $H_{eb}(T)$, and in addition the measured time evolution of the exchange bias field $H_{eb}$, i.e., $H_{eb}(t)$, we are able to calibrate the time evolution of the spin and lattice temperature $T(t)$ [26]. Real time temperature measurements on the multilayer stack and a comparison to the heat diffusion equation reveal that the spin temperature at the F/AF interface is elevated close to 100°C shortly after the pump pulse, thereby explaining the about 50% decrease of the exchange bias field within the first 10 ps [26]. Since the heat flow simulations fit only for pump-probe delay times larger than 20 ps, we deduce an interfacial spin-lattice relaxation time on the order of 20 ps, which agrees well with measured spin-lattice relaxation times of ferromagnetic multilayers by Guarisco *et al.* [27] and with the lower boundary measured by Vaterlaus *et al.* [17].

Upon the picosecond photoexcitation the exchange coupling is reduced and a nearly isotropic easy axis magnetization reversal behaviour is observed for both symmetry axes. The internal relaxation time τ for the slow recovery process of the exchange bias and coercive fields as well as of the zero-field susceptibility can be understood in terms of energy dissipation. The spin-lattice system is cooled by spin-lattice thermalization and heat flow from the metallic bilayer to the substrate or to a region outside the laser spot, setting the ultimate limit for the speed of recovery. Currently, the reduction of the exchange bias field is limited to about 48% of the initial value because, for larger pump pulse energies, irreversible processes occur which obstruct the stroboscopic measurement scheme used here. Since the samples are poly-crystalline, a distribution of different AF grain sizes is present. According to Takano *et al.* [28], large AF grains at the interface of an exchange bias system exhibit a smaller exchange coupling to the F spins across the interface. Upon laser excitation, these grains might already switch completely and will not relax to the original magnetization direction. In order to minimize the distribution width of the exchange bias field values, measurements of epitaxial exchange bias systems are planned.

The measured time evolution of the exchange bias shift field and the zero-field susceptibility can be interpreted as an anisotropy pulse field with a rise time on the order of the exciting pump pulse duration (see Fig. 3a and Fig 5a.). The spin temperature induced change of unidirectional exchange bias anisotropy leads to the observed unpinning effect. The effective pulse field is capable of triggering the measured magnetization oscillations. According to a general model first presented by van Kampen *et al.* [5], we can establish and follow different stages of the heat-induced precessional effect. For negative pump-probe delay times the magnetization points into its equilibrium direction defined by the exchange bias anisotropy. Upon sudden picosecond heating the exchange bias anisotropy of the F/AF bilayer changes dramatically altering the equilibrium orientation (see Fig. 3a and Fig. 5a). This leads to an initial "kick" on the F layer magnetization. Taking the fundamental Landau-Lifshitz and Gilbert equation (LLG) of motion

$$\frac{d\vec{M}}{dt} = -\gamma \, \vec{M} \times \vec{H}_{\text{eff}} - \frac{\gamma \, \alpha}{M_{\text{s}}} \, \vec{M} \times \left( \vec{M} \times \vec{H}_{\text{eff}} \right) \qquad (3)$$

and evaluating the expression for the precessional torque, we conclude that we are able to optically or thermally control the effective field $H_{\text{eff}}$. A time depending effective internal anisotropy pulse field exists which matches the measured time evolution of both the exchange bias field $H_{\text{eb}}(t)$ and the zero-field susceptibility $\chi(t)$. The initial kick on the magnetization of the ferromagnetic layer of the exchange bias bilayer leads to a torque ($\vec{M} \times \vec{H}_{\text{eff}} \neq 0$) on the magnetization, which triggers an initial precession around the new equilibrium orientation. Heat diffusion out of the F/AF interface and into the Cu buffer removes the excess heat, i.e., the original equilibrium orientation starts to be restored. However, due to its initial displacement, the F layer magnetization will continue to precess for hundreds of picoseconds.

Incorporating this excitation model in the LLG equation (Eq. (3)), e.g., in the effective field, we can deduce an approximative solution for the in-plane precessional angle Φ(t) upon a fast rising field pulse [29], with an amplitude β, the precession frequency $\omega_{\text{p}} = 2\pi f$ and a Landau-Lifshitz damping parameter $\lambda_{\text{LL}}$

$$\Phi(t) = \beta \cdot \sin(\omega_{\text{p}} t + \varphi) \cdot \exp(-t \cdot \lambda_{\text{LL}} / 2) \; . \qquad (4)$$

Equation (4) can be used to fit the measured precessional traces and to deduce the effective Gilbert damping parameter according to $\alpha = \lambda_{LL} / (|\gamma| 4\pi M_S)$, with $|\gamma|$ representing the gyromagnetic ratio. As the saturation magnetization we chose the pure Permalloy ($Ni_{81}Fe_{19}$) value of $4\pi M_S = 10.8$ kG. The fitting parameters $\beta$ and the phase $\varphi$ are optimized to reproduce the features of the individual precession traces. The precession frequency $\omega_p = 2\pi f$ and the effective damping contain the valuable physical information for us to infer, for instance, spin damping effects in the time domain. This analysis leads, for instance, to the exemplary fits in Figs. 3b and 4b (dotted lines overlapping the measured time-resolved trace) and in Fig. 5. A linear background has been assumed to mimic the reversible relaxation of the transient Kerr rotation to its initial value. The best fit yields an effective Gilbert damping parameter of $\alpha = 0.0475$ for the easy axis example case presented in Fig. 3b and Fig. 4b. This quite high effective damping parameter, i.e., dissipation rate can be motivated by an additional relaxation channel due to the F/AF interfacial exchange coupling [8], which is in qualitative agreement with FMR [7] and BLS linewidth measurements [30]. However, dephasing due to the granular nature of the investigated samples cannot be neglected and may also give rise to such high dissipation rates.

The possibility to control the strength of the internal pulse field for both symmetry directions of the investigated exchange coupled bilayer (see Fig. 4a and Fig. 5a) enables us to study the dependence of the effective Gilbert damping $\alpha$ on the angle of the precessional cone. The internal pulse field strength directly determines the initial magnetization deflection, and thus, the initial precessional cone angle. In the hard axis case, for instance, the precessional cone angle should monotonously decrease for applied static fields larger than 31 Oe (see dashed arrows in Fig. 5a), i.e., the precessional motion for both small and large deflection angles can be studied. The quasistatic reversal process in the hard axis case can be well described by coherent rotation between the two equilibrium orientations of the magnetization of the ferromagnetic layer. Coherent rotation can also be assumed as a fair first order approximation for the easy axis geometry reversal. Hence, we can calculate the initial angle for the precessional response for each applied static field value. Moreover, as described above, we can also deduce an effective Gilbert damping for each recorded transient Kerr trace according to Eq. (4). Figure 7 summarizes the analysis of the dependence of the effective Gilbert damping on the extracted precessional cone angle, which covers a range from 10 degree up to about 135 degree. Full squares represent the data deduced from the hard axis measurements, which cover the small and intermediate angle as well as the onset of large angle precessional motion. The open square was deduced from the maximum amplitude precession for the easy axis case, which exemplarily accounts for a large angle precession. Within the error bars (dashed lines represent 2σ intervals) the effective Gilbert parameter stays constant, indicating that there are no or only minor nonlinear effects present in the measured intermediate angle range for the investigated bilayer system. A roughly 5 % increase of the effective Gilbert damping is observed for the large angle easy axis precession compared to the values found in the hard axis geometry. Again, this minor increase suggests that both a large and a small angle precessional magnetization response to an internal pulse field excitation can be modeled with reasonable accuracy in terms of an effective Landau-Lifshitz and Gilbert damping. These results are in good agreement with those reported by Nibarger *et al*. [15]. Presumably, nonlinear effects only play a crucial role in large angle precession or full switching events.

From an application's point of view, the measured coherent precessional motion of the magnetization of the ferromagnetic layer suggests the possibility of inducing a full coherent magnetization rotation or even complete switching of the F layer depending on the timing and especially the width and strength of the internal anisotropy pulse field. This, for instance, may be used for ultrafast heat-assisted recording in exchange biased systems on a picosecond timescale with temperatures involved far below typical Curie temperatures of ferromagnets.

In summary, it has been shown that picosecond short optical pulses can affect the magnetization reversal behavior of a NiFe/FeMn exchange bias system on a picosecond time scale. For easy and hard axis magnetization reversal, the same internal relaxation time constants upon photoexcitation are observed, indicating that the origin of this effect relies on the optical control of the interfacial exchange coupling. The measured time evolution of both the exchange bias field and the zero-field susceptibility can be understood in terms of a thermally controlled internal anisotropy pulse field which enables us to trigger and all-optically probe coherent precessional magnetization dynamics in an exchange coupled bilayer on the picosecond timescale. Both the direction of the initial magnetization deflection and the strength of the internal pulse field can even be controlled by the absorbed photons. The extracted effective Gilbert damping reveals no significant dependence on the precession angle, indicating that both large angle as well as small angle precessional magnetization dynamics due to an internal pulsed field excitation can be modeled with reasonable accuracy in terms of an effective Gilbert damping.

**Acknowledgments**

The authors would like to thank C. Bayer, B. Beschoten, S. Blomeier and K. O´Grady for valuable discussions. One of the authors (M.C.W.) would like to acknowledge support by the Graduiertenkolleg 792 "Nichtlineare Optik und Ultrakurzzeitphysik" of the Deutsche Forschungsgemeinschaft. The work is supported in part by the European Communities Human Potential programs under contract No. HPRN-CT-2002-00318 ULTRASWITCH and HPRN-CT-2002-00296 NEXBIAS.

**Figure captions**

**Fig. 1:** Time dependence of the exchange bias field $H_{eb}(t)$. A model fit to Eq. (1) with the parameters $m = 0.48$ and $\tau = 160$ ps is shown by a full line. The open squares show the measured time evolution of the easy axis coercivity $H_c(t)$.

**Fig. 2:** Time dependence of the zero-field susceptibility $\chi(t)$ of a hard axis magnetization reversal loop around 0 ps delay time together with the measured intensity autocorrelation of the pump pulse. The dashed-dotted line represents a fit to Eq. (2) with a time constant of $\tau^* = 160$ ps and a modified modulation strength $m^* = 13.75$.

**Fig. 3:** (a) Determination of the magnitude of the internal pulse field motivated by two easy axis loops with (dashed line) and without (solid line) pump induced exchange decoupling, which can alter the equilibrium orientation of the magnetization. The thick arrow indicates and quantifies the magnetization deflection from the equilibrium direction. The small grey arrow highlights the slight in-plane rotation of the magnetization of the ferromagnetic layer for the marked case of $H_{stat} = 125$ Oe. (b) Magnetization oscillations for maximum internal pulse field strength ($H_{stat} = 125$ Oe) with a fit to Eq. (4).

**Fig. 4:** (a) Control of the strength of the internal pulse field. The open and full circles indicate different static field values, thus, different strengths of the internal pulse field. (b) Real time Kerr traces for applied fields values as indicated in (a). A maximum oscillation amplitude is found for the case of $H_{stat} = 125$ Oe (see also Fig. 3b). Exemplary fits to Eq. (4) are shown by dotted lines. Traces are offset for clarity.

**Fig. 5:** (a) Scheme of the anisotropy pulse fields for the hard axis case. The dashed and full arrows indicate different static field values, thus, different magnitudes of the internal pulse field. The direction of the arrows displays the onset of magnetization deflection. (b) Synopsis of the real time Kerr traces for applied field values as indicated in (a). The grey arrows highlight the field dependent oscillation period. Exemplary fits to Eq. (4) are presented by dotted lines. Traces are offset for clarity.

**Fig. 6:** Field dependence of the hard axis oscillation frequencies for positive applied fields. A square root Kittel like behavior can be deduced, as indicated by the dashed line.

**Fig. 7:** Effective Gilbert damping for the hard axis (full squares) and easy axis (open square) precession as a function of the precessional angle. Assuming coherent rotation the cone angle of the precession could be determined, thus, reflecting the strength of the internal pulse field. Note the break in the abscissa.

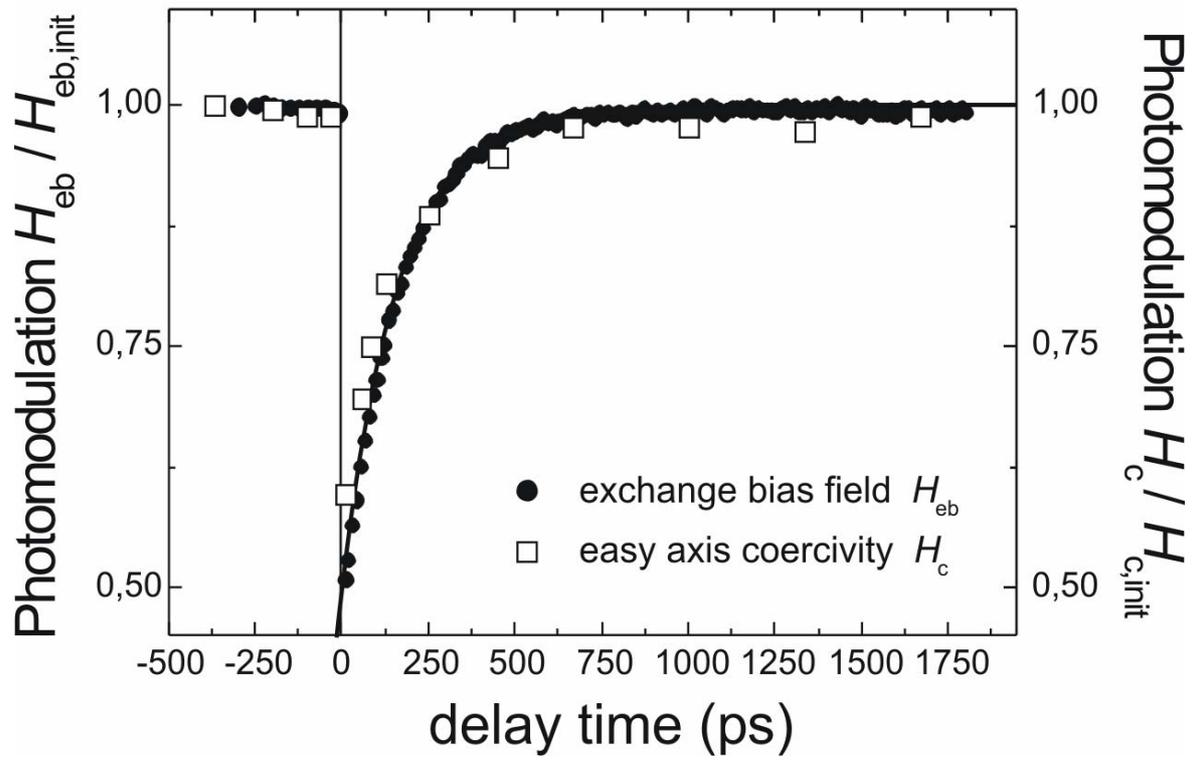

Figure 1

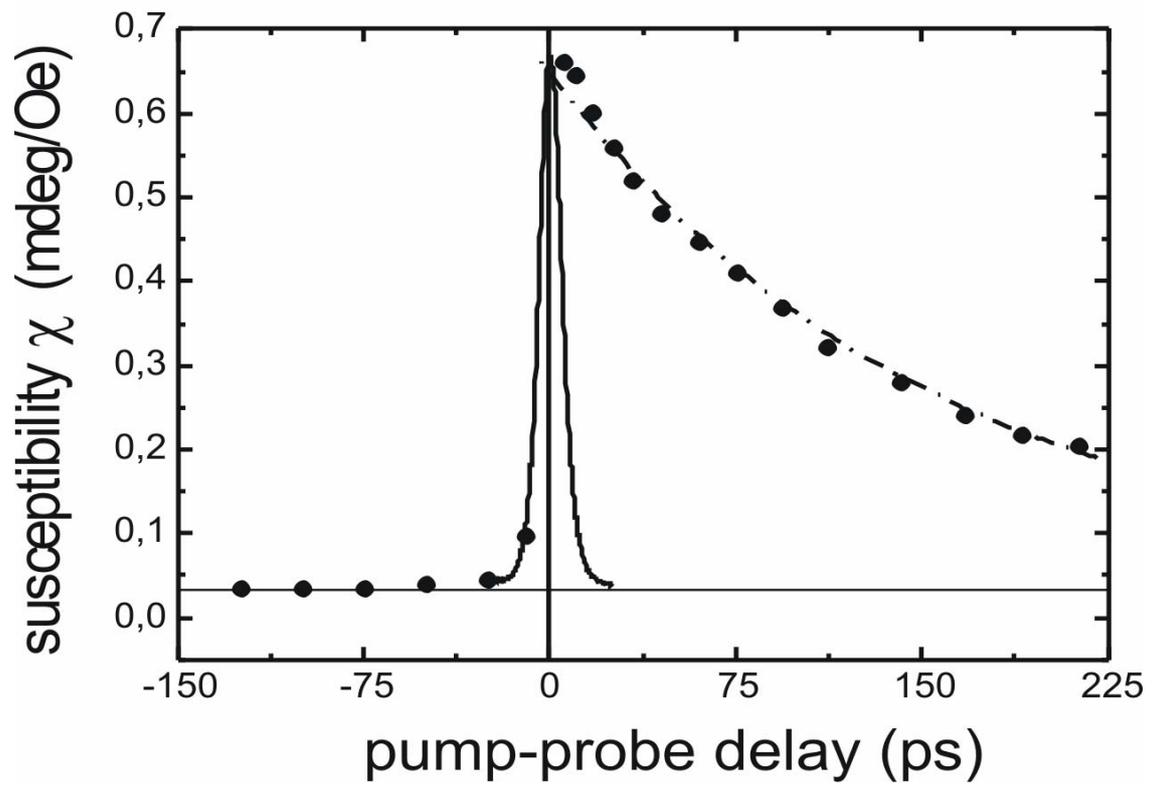

Figure 2

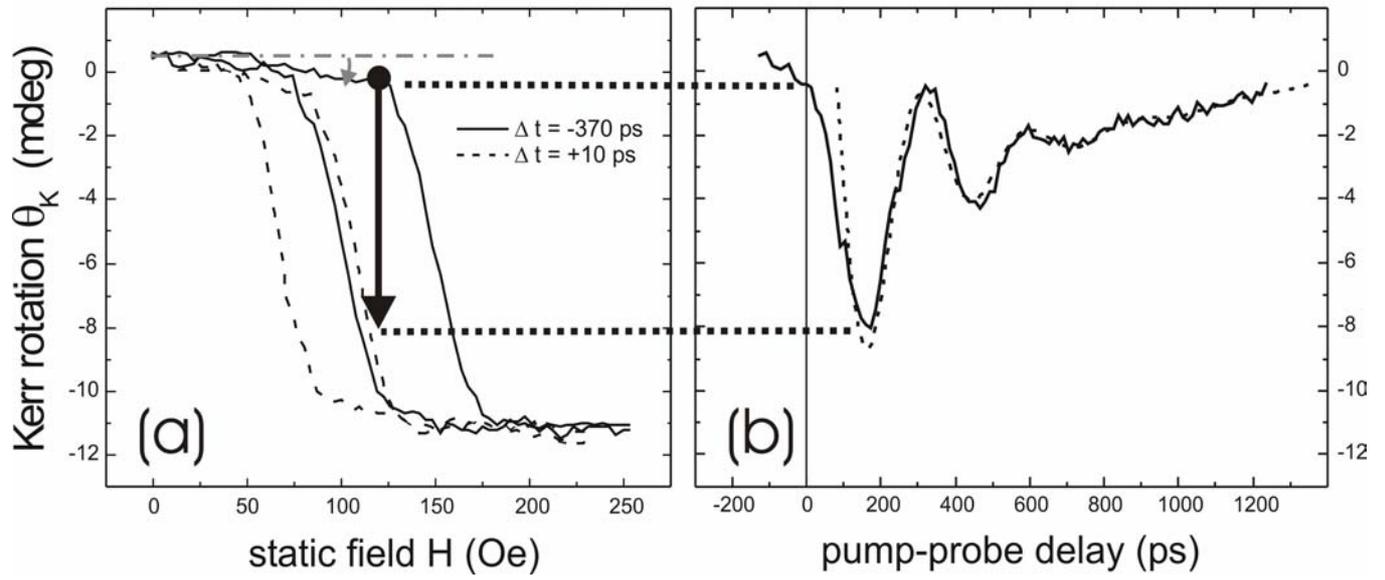

Figure 3

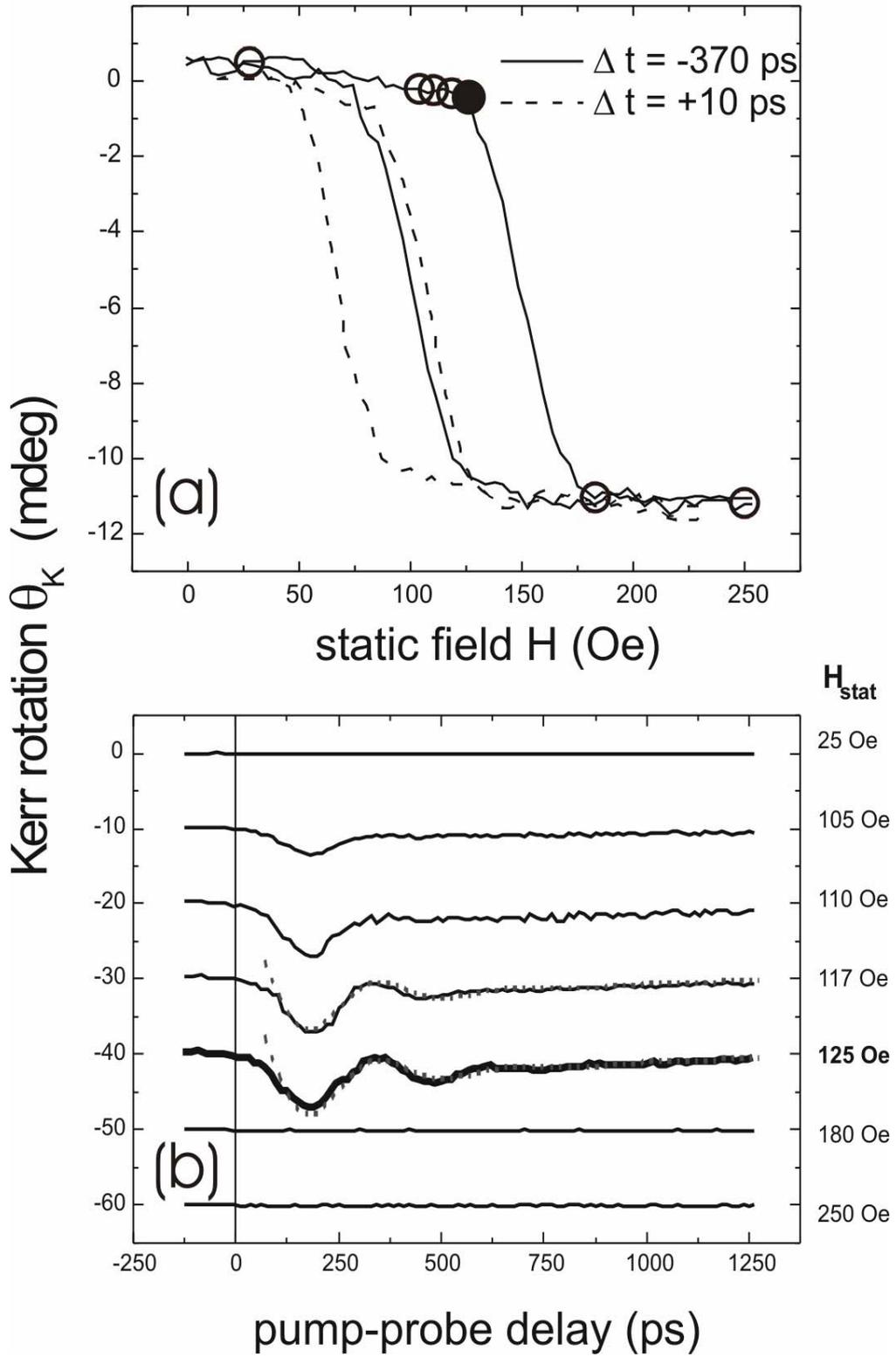

Figure 4

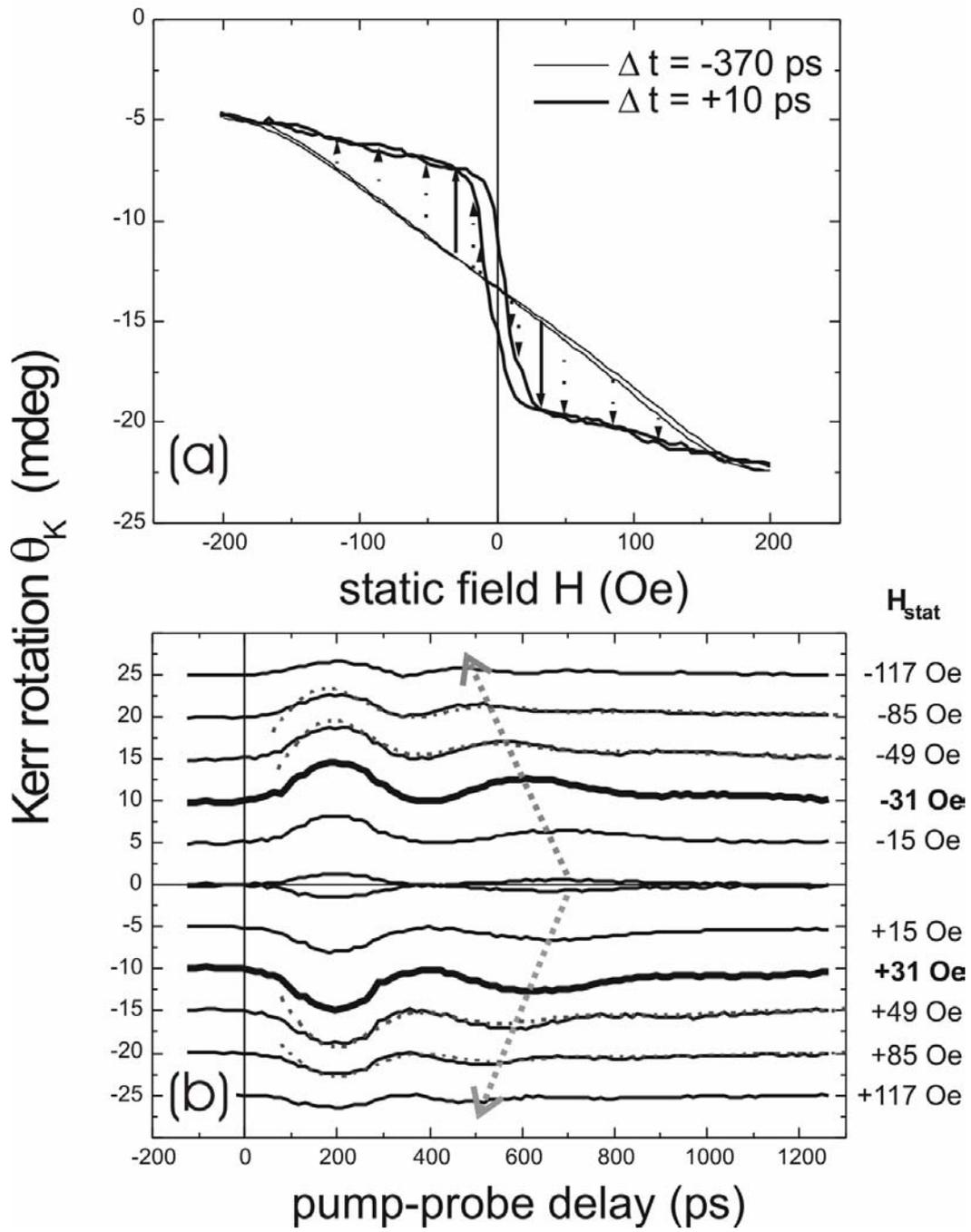

Figure 5

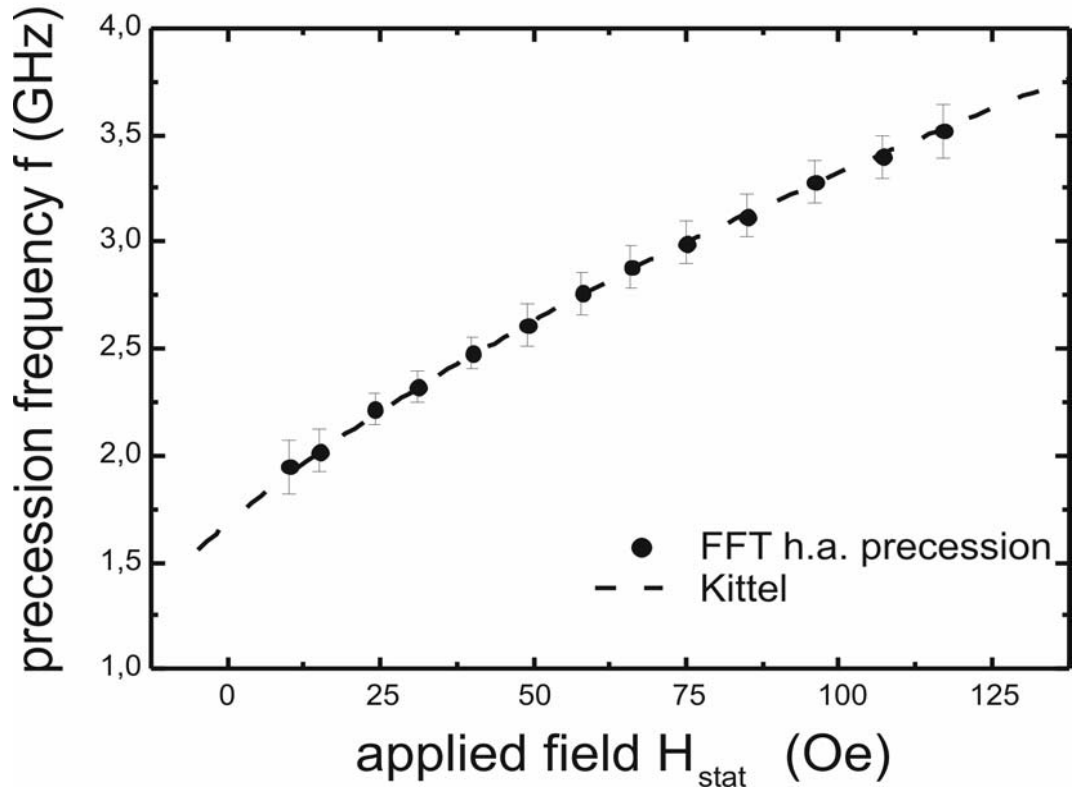

Figure 6

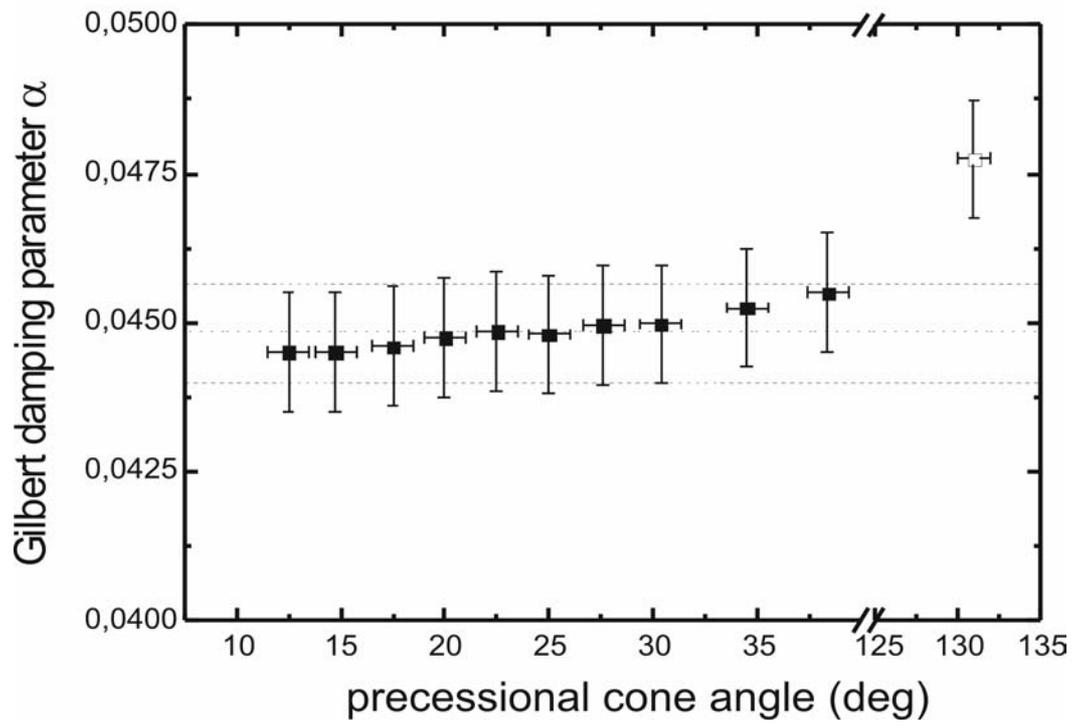

Figure 7